\documentclass{elsart}    
\usepackage{amsmath,amsfonts,amssymb}
\usepackage[sort&compress,square,comma]{natbib}
\usepackage{graphicx} 
\usepackage{color} 
\journal{Physica A}
\begin{document}
\begin{frontmatter}   \title{Exact  solution   of   a  one-dimensional
Boltzmann    equation     for    a    granular     tracer    particle}
\author[warsaw]{J.     Piasecki}     \author[pittsburgh]{J.    Talbot}
\author[paris]{P. Viot}

\address[warsaw]{Institute of Theoretical Physics, University of Warsaw, Ho\.za 69, 00-681 Warsaw, Poland} 

\address[pittsburgh]{Department   of  Chemistry  and  Biochemistry,
Duquesne University, Pittsburgh, PA 15282-1530, USA}
 \address[paris]{Laboratoire de Physique
Th\'eorique de la Mati\`ere Condens\'ee, Universit\'e Pierre et Marie Curie, 4, place
Jussieu, 75252 Paris Cedex, 05 France}
\begin{abstract} We consider a  one-dimensional system consisting of a
granular tracer   particle  of mass   $M$  in a  bath  of  thermalized
particles each of  mass $m$.  When the mass  ratio, $M/m$, is equal to
the coefficient   of  restitution, $\alpha$,   the system   maps to    a  a
one-dimensional elastic gas.  In  this case, Boltzmann equation can be
solved exactly.  We      also obtain expressions   for    the velocity
autocorrelation function   and the  diffusion  coefficient.  Numerical
simulations of the Boltzmann equation are performed for $M/m\neq \alpha$ where
no  analytical solution is available.  It  appears  that the dynamical
features  remain qualitatively similar to   those found in the exactly
solvable case.
\end{abstract}

\end{frontmatter}

\section{Introduction}  Granular  gases  consist of  particles that
undergo   dissipative   collisions,   making   them  a   paradigm   of
non-equilibrium  statistical mechanics \cite{MP99,BT02}.   The absence
of  microscopic reversibility  (in each  collision, kinetic  energy is
lost) and the  contraction of the phase space  volume lead to specific
properties  which differ from  elastically colliding  thermal systems.
These  include non-gaussian  statistics,  modified hydrodynamics,  and
absence of equipartition \cite{MP99,BT02,WJM03,FM02,WP02,KT03}.

Compared to  equilibrium, our knowledge  of nonequilibrium statistical
mechanics is  still incomplete, and  exact results are rare.   In this
paper, we consider  a one-dimensional model of a  tracer particle that
undergoes dissipative collisions with particles of a thermalized bath.
The  kinetic description is  provided by  the Boltzmann  equation.  We
show that when the mass ratio of the tracer to a bath particle, $M/m$,
is equal  to the coefficient  of restitution $\alpha$, there  exists a
mapping with  the Boltzmann equation  of identical elastic  hard rods,
the   solution  of  which   was  derived   almost  thirty   years  ago
\cite{R78}. We  propose here  a simpler way  of solving  the Boltzmann
equation and we discuss the results  in the context of our model, i.e.
a granular particle in a thermalized bath.  In addition, we obtain the
velocity   autocorrelation   function  as   well   as  the   diffusion
coefficient.   Numerical simulations  of the  Boltzmann  equation show
that the  qualitative kinetic features  are not strongly  dependent on
the value of  the coefficient of restitution, implying  that the exact
solution  provides  the  characteristics   of  the  system  also  when
$M/m\neq\alpha$, where an analytic solution is not available.

\section{The  model} We  consider a  one-dimensional gas  of identical
hard rods  of mass $m$ into which  one inserts a granular  hard rod of
mass  $M$\cite{exactptv}.  Particles have  hard core  interactions and
collisions between bath particles are elastic. But, collisions between
the tracer particle  and the bath particles are  assumed inelastic and
characterized  by a  coefficient of  restitution, $\alpha\leq  1$.  In
addition, we assume that  a stationary, Gaussian velocity distribution
is imposed on the bath  particles, compensating for the kinetic energy
lost during collisions with the tracer particle.

For each
collision, momentum  is conserved  so that for  a collision  between a
bath particle and the tracer particle, one has
\begin{equation}\label{eq:1}
MV^*+m v^*=M V+m v
\end{equation}
where the upper-case  velocity corresponds to  the tracer particle and
the lower-case velocity to the bath  particle, and the asterisks denote
post-collisional  quantities. At the moment of impact, the relative velocity
changes sign and shrinks (for $\alpha < 1$) according to the collision rule
\begin{equation}\label{eq:2}
 V^*- v^*=-\alpha( V- v)
\end{equation}
By combining  Eqs.(\ref{eq:1})  and (\ref{eq:2}),   the
velocities of colliding particles after  collision are given by
\begin{align}\label{eq:3}
 V^*=\frac{(\mu-\alpha) V+(1+\alpha) v }{1+\mu}\\\label{eq:4}
 v^*=\frac{(1+\alpha) V+(\mu^{-1}-\alpha) v }{1+\mu^{-1}}
\end{align}
where $\mu=M/m$.

Note that  the tracer  mass can  be also  taken as
equal  to  the  fluid   particle  mass  by  introducing  an  effective
restitution coefficient\cite{PVTW06}.

The precollisional velocities  $ V^{**}$ and $ v^{**}$,  corresponding
to the inverse collision
\[  (V^{**},v^{**}) \to (V,v) ,\] 
 can  be expressed in  terms of $ V$ and   $ v$ by replacing in
 Eqs.(\ref{eq:3}) and  (\ref{eq:4}) the coefficient  $\alpha$ by  $\alpha^{-1}$.

\section{Structure of the Boltzmann equation at $\alpha = M/m$}

The kinetic  description of this  model is  provided by  the Boltzmann
equation according to which the probability distribution of the tracer
particle evolves as
\begin{align}\label{eq:5}
&\left( \frac{\partial }{\partial t}+ V\frac{\partial }{\partial R }\right) f( R, V,t)=
\nonumber\\
&=\rho \int_{-\infty}^{+\infty}dc| V- c|\left[ \alpha^{-2}f(R, V^{**},t)\phi( c^{**})-f( R, V,t) \phi(c)
\right]
\end{align}

where $R$ is the position of the tracer particle, $\phi(c)$ denotes the
equilibrium Maxwell distribution of the bath particles, and $\rho $
their number density.  Note that Eq. (\ref{eq:5}), although linear in
the case of a tracer particle, has not yet been solved exacty.  Exact
solutions have, however, been found in the framework of the inelastic
Maxwell model where the collision frequency, which occurs in the
Boltzmann collisional term, is assumed independent of the relative
velocity \cite{BK03,EB02}.

The first objective of this paper is to show that an exact solution of
Eq. (\ref{eq:5})  can be obtained when the  coefficient of restitution
equals   the  mass   ratio  of   the  tracer   to  a   bath  particle,
$\alpha=\mu=M/m$.   In  this  case,  the Boltzmann  equation  for  the
distribution  of  the  granular   tracer  particle  turns  out  to  be
mathematically identical to the Boltzmann equation describing a tagged
elastic  particle  in a  thermalized  bath  of mechanically  identical
particles. This fact, shown  below, opens the way to an
exact  solution.    However,  the  physical   situations  are  totally
different:  evolution of  a  dissipative system  towards a  stationary
state  in one  case and  evolution  of a  conservative system  towards
thermal equilibrium in the other.

The differences between the two systems on the level of the
microscopic dynamics are illustrated by the collision rules.  For the
elastic system with identical masses, a binary collision leads to the
exchange of particle velocities, $V^*=c$ and $c^*=V$.  Moreover, for
the inverse collision one finds $V^{**}=c$ and $c^{**}=V$, which is
related to the microscopic reversibility.  For the granular tracer
particle of mass $M=\alpha m$ in a thermalized bath of particles of mass
$m$, the collision rules are $c^*=\alpha V+(1-\alpha)c$ and $V^*=c$, whereas
for the precollisional velocities corresponding to the inverse
collision one finds $c^{**}=V$ and $V^{**}=c+(1-\alpha)V$.  Despite the
differences, in the gain term of the Boltzmann collision operator, in
both cases, the precollisional velocity of the bath particle equals
$V$, and is thus independent of $c$.  And this common feature results in
the simplification that allows the equation to be solved, as can be seen
below.

For    the  sake of  simplicity, we   first   consider the 
Boltzmann equation for spatially homogeneous systems. The general solution 
for inhomogeneous initial conditions can be derived in a very similar way.  
The corresponding, rather tedious, calculations are given in  Appendix A.

The Boltzmann equation for homogeneous states takes the form 
\begin{align}
\label{eq:6}
&\frac{\partial      f}{\partial t}(    V,t)=\rho \int_{-\infty}^{+\infty}dc|    V-   c|\left[\alpha^{-2}
f\left(\frac{(\mu-\alpha^{-1})         V+(1+\alpha^{-1})         c          }{1+\mu}
,t\right)\right.\nonumber\\                  &\left. \times\phi\left(\frac{(1+\alpha^{-1})
V+(\mu^{-1}-\alpha^{-1}) c }{1+\mu^{-1}}\right)-f( V,t) \phi(c)\right].
\end{align}

Putting $\mu = \alpha$ in the  gain term of
Eq.(\ref{eq:6}) yields
\begin{equation}
\label{eq:7}
\left( \frac{\partial }{\partial t}+\nu(V)\right) f(V,t)=\rho \phi(V) \int_{-\infty}^{+\infty}dc| V- c|\alpha^{-2}
f\left(\frac{(\alpha-1) V}{\alpha}+\frac{c}{\alpha} ,t\right)
\end{equation}
where $\nu(V)$, the collision rate of the tracer,  is given by
\begin{equation}
\label{eq:8}
\nu(V) = \rho \int_{-\infty}^{+\infty}dc|V-c|\phi(c) 
\end{equation}

By using a change of the integration variable, $u=c/ \alpha+(\alpha-1)V/ \alpha$,
in the right-hand side of Eq.(\ref{eq:7}), one finally obtains
\begin{equation}
\label{eq:9}
\left( \frac{\partial }{\partial t}+\nu(V)\right) f(V,t)=\rho \phi(V) \int_{-\infty}^{+\infty}du| V- u|f(u ,t),
\end{equation}
which coincides exactly with the Boltzmann equation for elastically colliding identical particles of mass $m$, solved by R\'esibois \cite{R78}.

  Before determining the time evolution of $f(V,t)$, we first focus
on  the stationary  state $f^{st}(V)$ which satisfies the equation obtained by dropping   the time  derivative in
Eq.(\ref{eq:9}):
\begin{equation}
\label{eq:10}
f^{st}(V) \int_{-\infty}^{+\infty}dc| V- c|\phi(c)= \phi(V) \int_{-\infty}^{+\infty}dc| V- c|f^{st}(c),
\end{equation}
The solution of (\ref{eq:10}) is given by
\begin{equation}
\label{eq:11}
f^{st}(V)=\phi(V)=\sqrt{\frac{m}{2\pi T}}\exp\left(-\frac{mV^2}{2T}\right)=\sqrt{\frac{M}{2\pi\alpha T}}
\exp\left(-\frac{MV^2}{2\alpha T }\right)
\end{equation}
where $T$ is temperature of the bath.  Therefore,  the stationary velocity distribution is Maxwellian,
but with a temperature $\alpha T$ smaller than the  bath temperature.  This
result is a specific case of a previously obtained general formula valid
for any mass ratio in any dimension \cite{MP99}. 

From Eq. (\ref{eq:8}) we get an explicit expression for the collision frequency of the granular particle
\begin{equation}
\nu(V)=\rho    \left[\sqrt{\frac{2\alpha  T}{\pi   M   }}\exp\left(-\frac{MV^2}{2\alpha
T}\right)+V \,erf\left(\sqrt{\frac{M}{2\alpha T}}V\right)\right]
\end{equation}
where $erf(x)$ is the error function.

In the following, we use dimensionless variables: the units of length
and time are $1/ \rho$ and $ \sqrt{m}/\sqrt{T}\rho$, respectively. Also,
for the sake of simplicity, we use lower-case variables since no
confusion with the variables associated with the bath particles is
possible.
  
\section{Exact Homogeneous solution}
We first show that the integro-differential equation (\ref{eq:9})  can be replaced by
a differential  equation.   To this end we  introduce  an auxiliary function $G(u,t)$
\begin{equation}
\label{eq:12}
G(u,t)=\int_{-\infty}^{+\infty}dc|u-c|f(c,t)
\end{equation}
 satisfying the relation
\begin{equation}
\label{eq:13}
\frac{\partial^2G}{\partial u^2}(u,t)=2f(u,t).
\end{equation}
Therefore, the Boltzmann equation,  Eq.(\ref{eq:9}), when re-expressed in
terms of $G(u,t)$, yields
\begin{equation}\label{eq:14}
\frac{\partial^3G}{\partial t \partial u^2}(u,t)+\nu(u)\frac{\partial^2G}{\partial u^2}(u,t)=
\frac{d^2\nu }{d u^2}(u)G(u,t)
\end{equation}

Let us introduce the time Laplace transform
\begin{equation}\label{eq:15}
\tilde{G}(u,z)=\int_0^{+\infty}dt \, G(u,t)e^{-zt},
\end{equation}
which when applied to Eq.(\ref{eq:14}) yields 
\begin{equation}\label{eq:16}
[z+\nu(u)]\frac{\partial^2 \tilde{G}}{\partial u^2}(u,z)-\frac{d^2\nu (u)}{d u^2}\tilde{G}(u,z)=\frac{\partial^2 G}{\partial u^2}(u,0)
\end{equation}

By using the identity
\begin{equation}\label{eq:17}
\nu(u)\frac{\partial^2\tilde{G}}{\partial u^2}(u,z)-\frac{d^2\nu}{d u^2}(u)\tilde{G}(u,z)=
\frac{\partial}{\partial u}\left[\nu (u)
\frac{\partial \tilde{G}}{\partial u}(u,z)-\frac{d\nu }{d u}(u)\tilde{G}(u,z)\right],
\end{equation}
we can integrate Eq.(\ref{eq:16}) over  the velocity from $-\infty$ to $u$ finding
\begin{equation}
\label{eq:18}
[z+\nu(u)]\frac{\partial\tilde{G}}{ \partial u}(u,z)-\frac{d\nu}{d u}(u)\tilde{G}(u,z)=\frac{\partial G}{\partial u}(u,0)+\tilde{B}(z)
\end{equation}
where  $\tilde{B}(z)$ is a function that  can be determined by taking in (\ref{eq:18}) the limit $u\to \infty$. 
According to the definitions (\ref{eq:12}),(\ref{eq:15})  in the region of large velocities  $\tilde{G}(u,z)$ is asymptotically given by
\begin{equation}
\label{eq:19}
\tilde{G}(u,z)=\frac{|u|}{z}-sgn(u)<\tilde{v}(z)>
\end{equation}
where  $<\tilde{v}(z)>$ is the  Laplace  transform of the mean velocity
\begin{align}
<\tilde{v}(z)>=\int_{-\infty}^{+\infty}du u \tilde{f}(u,z).
\end{align}
Using Eq.(\ref{eq:19}) one can readily check that Eq.(\ref{eq:18}) considered in the limit $|u|\to\infty$ 
reduces to the equality $\tilde{B}(z)=<\tilde{v}(z)>$.  

Introducing now the function 
\begin{equation}
\label{eq:20}
\tilde{H}(u,z)=\frac{\tilde{G}(u,z)}{[z+\nu(u)]}
\end{equation}
we rewrite Eq.(\ref{eq:18}) as
\begin{equation}
\label{eq:21}
\frac{\partial \tilde{H}}{\partial u}(u,z)=\frac{1}{[z+\nu(u)]^2}\left[ \frac{\partial G}{\partial u}(u,0)+<\tilde{v}(z)>\right]
\end{equation}

From the   asymptotic formula   Eq.(\ref{eq:19})  combined  with   the
definition (\ref{eq:20}), it follows that
\begin{equation}\label{eq:22}
\lim_{u\to \pm \infty}\tilde{H}(u,z)= \frac{1}{z}.
\end{equation}
Using this result we can integrate  Eq.(\ref{eq:21}) over the velocity space obtaining  an explicit expression  for  the  mean velocity 
\begin{equation}
\label{eq:23}
<\tilde{v}(z)>=-\int_{-\infty}^{+\infty}\frac{du}{[z+\nu(u)]^2}\frac{\partial G}{\partial u}(u,0)\left(\int_{-\infty}^{+\infty}\frac{du}{[z+\nu(u)]^2}\right)^{-1}
\end{equation}

The initial condition appears in Eq. (\ref{eq:23}) through the derivative (see Eq. (\ref{eq:12}))
\begin{equation}
\label{eq:233}
 \frac{\partial G}{\partial u}(u,0) = \int_{-\infty}^{u}dc\, f(c,0)-\int_{u}^{+\infty}dc\, f(c,0) 
\end{equation}
 It  is  worth  noting that  we could calculate the first  moment of the probability 
distribution $<\tilde{v}(z)>$ without yet knowing the complete solution of the Boltzmann equation.
If $f(u,0)$  is  an even  function of
velocity, then the derivative (Eq. \ref{eq:233})  is an odd  function, and  
from Eq. (\ref{eq:23}) we get $<\tilde{v}(z)>=0$, and hence the mean velocity $<v(t)>=0$, for any $t>0$.

The  differential equation, Eq.(\ref{eq:21})  can be easily integrated
with the use of  the boundary condition, Eq.(\ref{eq:22}).  The
relation Eq. (\ref{eq:20})  between $\tilde{G}(u,z)$   and  $\tilde{H}(u,z)$ yields  then the complete solution 
of Eq.(\ref{eq:18})
\begin{equation}
\label{eq:24}
\tilde{G}(u,z)=[z+\nu(u)]\left[\int_{-\infty}^udw\frac{1}{[z+\nu(w)]^2}
\left(\frac{\partial G}{\partial w}(w,0)+<\tilde{v}(z)>\right)+\frac{1}{z}\right]
\end{equation}

Differentiating  Eq.(\ref{eq:24}) twice with
respect  to $u$ we find the  Laplace transform $\tilde{f}(u,z)$ of the
solution of the Boltzmann equation
\begin{align}
\label{eq:25}
\tilde{f}(u,z)&=\phi(u)\left[\frac{1}{z}+\int_{-\infty}^udw\frac{1}{[z+\nu(w)]^2}
\left(\frac{\partial G}{\partial w}(w,0)+<\tilde{v}(z)>\right)\right]
\nonumber\\&+\frac{1}{[z+\nu(u)]}f(u,0).
\end{align}

Eq.(\ref{eq:25})  when integrated over the  velocity yields
 $\int dv \tilde{f}(u,z)=1/z$, a result consistent with the conservation of the probability.

The use of the  identity  
\[ \frac{1}{[z+\nu(u)]}+\int_{-\infty}^udw \frac{1}{[z+\nu(w)]^2}\frac{d\nu (w)}{dw}=0 \]   
in Eq.(\ref{eq:25}) allows us to perform   the inverse Laplace  transform term
by term. In this way we determine the structure of the time dependent solution of the 
Boltzmann equation
\begin{align}
\label{eq:26}
f(u,t)&=\phi(u)+[f(u,0)-\phi(u)]e^{-t\nu(u)}\nonumber\\
&+\phi(u)\left[\int_{-\infty}^udw\left(\left[\frac{\partial G}{\partial w}(w,0)-
\frac{d\nu(w)}{dw}\right]te^{-t\nu(w)}+A(w,t)\right)\right]
\end{align}

where $A(w,t)$ is the inverse Laplace transform of
\begin{equation}
\tilde{A}(w,z)=\frac{<\tilde{v}(z)>}{[z+\nu(w)]^2}.
\end{equation}

Note that when the initial distribution $f(u,0)$ is  an  even function of $u$
(and thus $<\tilde{v}(z)>=0$),  $A(w,t)$ vanishes.

\begin{figure}[t]
\begin{center}
\resizebox{11cm}{!}{\includegraphics{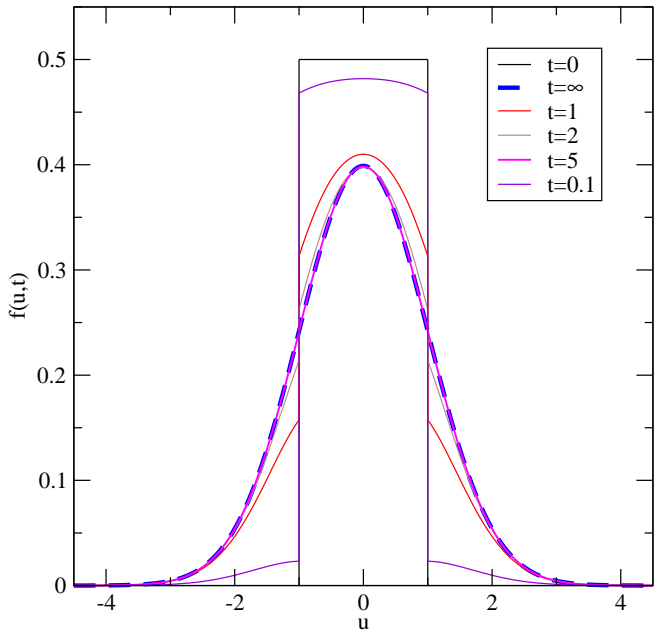}}
\caption{Velocity distribution function $f(u,t)$ versus $u$ at different times $t=0.1,1,2,5,\infty$.}
\end{center}
\label{fig:1}
\end{figure}

The first  term  on the right-hand side of
Eq.(\ref{eq:26}) represents the   stationary solution, whereas  the
second  term  describes the exponential decay (with relaxation time 
 $1/ \nu(u)$) of the  initial  deviation from  the
stationary distribution.  For high velocities the collision frequency $\nu(u)$ goes
as $|u|$ so that the associated  relaxation time approaches $0$ as
$ |u|^{-1}$. Consequently, the high energy tails of the distribution function
converge  faster towards the stationary  state than  the rest of the
distribution. This phenomenon is  illustrated in  Fig.\ref{fig:1} which shows the
evolution of the  velocity distribution starting from a 
``gate''  function. The initial discontinuities in velocity space are present at any finite
time. Their position remains fixed, but their magnitude decreases with time.
Finally, we note that the last term  on the right-hand side of Eq.(\ref{eq:26}) vanishes both for
$t\to 0$, and for  $t\to\infty$.

\section{Velocity autocorrelation function and  diffusion coefficient}
\label{sec:corr-funct-transp}
The velocity autocorrelation function is defined  by 
\begin{equation}
C(t)=<v(t)v(0)>
\end{equation}
where the brackets  denote an average over  the  stationary Maxwell distribution  (Eq. \ref{eq:11}). 
We can calculate $C(t)$ exactly using the solution (\ref{eq:25}). 
Specifically,
\begin{equation}
C(t)=\int_{-\infty }^{+\infty}v F(v,t)dv
\end{equation}
where $F(v,t)$ is the solution of the Boltzmann equation corresponding to the  initial condition
\begin{equation}
\label{eq:27}
F(v,0)=v\phi(v)
\end{equation}

The Laplace transform of $C(t)$
\begin{equation}
\label{eq:28}
\tilde{C}(z)=\int_{-\infty }^{+\infty}v \tilde{F}(v,z)dv = <\tilde{V}(z)>
\end{equation}
is thus equal to the Laplace transform of the mean velocity $<V(t)>$  corresponding to the
solution $\tilde{F}(v,z)$ (see Eq.(\ref{eq:23})).
Note that the initial condition (\ref{eq:27}) does not represent a probability distribution. 
One finds in this specific case the relation
\begin{equation}
\label{eq:29}
\frac{\partial G}{\partial u}(u,0)=u\frac{d\nu(u)}{d u}-\nu(u).
\end{equation}
Inserting Eq.(\ref{eq:29})  into Eq.(\ref{eq:23}) after integration by parts we get
\begin{equation}
\label{eq:30}
\tilde{C}(z) = <\tilde{V}(z)>=2\left(\int_{-\infty}^{+\infty}\frac{du}{[z+\nu(u)]^2}\right)^{-1}-z.
\end{equation}
Eq.(\ref{eq:30}) is an exact formula for the velocity autocorrelation function. The above derivation is by far simpler than that given in \cite{R78}.
\begin{figure}[t]
\begin{center}
\resizebox{11cm}{!}{\includegraphics{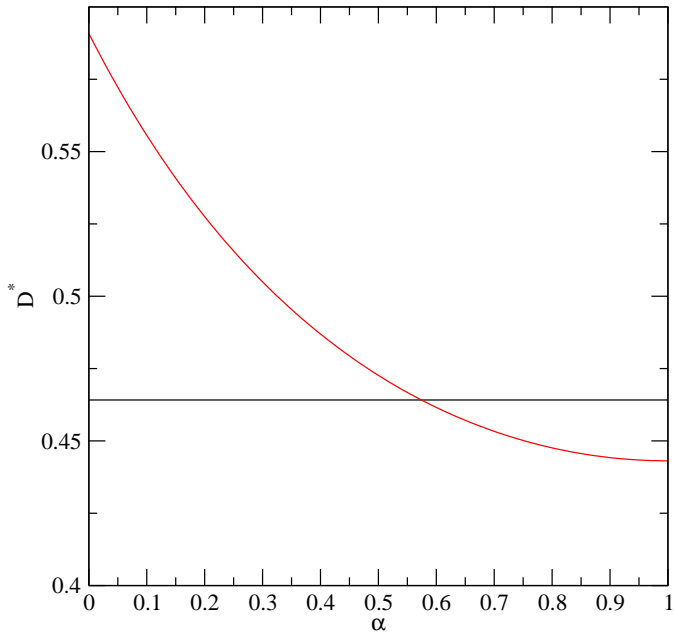}}

\caption{Reduced diffusion coefficient as a function of  the coefficient of restitution: 
the horizontal  line  corresponds to  the  exact solution  of the  Boltzmann
equation and the curve to the first Sonine approximation.}\label{fig:2}

\end{center}
\end{figure}

The dimensionless diffusion  coefficient $D^*=\tilde{C}(0)$ can be now deduced by taking the limit $z\to 0$  in Eq.(\ref{eq:30}). We find
\begin{equation}
\label{eq:31}
D^*=\left[ \int_{0}^{+\infty}\frac{du}{\nu(u)^2} \right]^{-1}
\end{equation}
Restoring the   physical dimensions of the
variables yields for the diffusion coefficient $D$ the formula
\begin{equation}
\label{eq:32}
D=\frac{1}{\rho}\sqrt{\frac{\alpha T}{M}}D^*
\end{equation}

The numerical  value of $D^*$  is $0.464139...$.  It is worth noting that
$D$ is proportional to the square root of the coefficient of restitution and hence
vanishes when $\alpha\to 0$.

The diffusion coefficient for a tracer particle has also been  obtained
approximately from a  normal Chapmann-Enskog
solution of the Boltzmann equation \cite{BRM05,GD02} yielding
\begin{equation}
\label{eq:33}
D^*=\frac{\sqrt{\pi}}{3+2\alpha-\alpha^2}.
\end{equation}
Figure  \ref{fig:2}   shows  the  reduced   (dimensionless)  diffusion
coefficient $D^*$ versus the coefficient of restitution $\alpha$.  

We note that the diffusion coefficient for elastic ($\alpha=1$) hard rods
in one-dimension is known exactly: $D^*=1/\sqrt{2\pi}\simeq
0.3989..$\cite{RdL77}.  Probably fortuitously, at $\alpha =1$ the lowest
order Sonine approximation to the solution of the Boltzmann equation
gives a result for $D^*$ that is closer to this exact result than the
prediction of the Boltzmann equation itself.  When $\alpha<1$, the reduced
diffusion coefficient within the Sonine approximation increases when
$\alpha$ decreases, whereas the result (Eq.  \ref{eq:31}) provided by the
Boltzmann equation is independent of $\alpha$.  The two curves cross at
$\alpha=0.574$. In both cases, of course, the full diffusion coefficient
$D$ vanishes when $\alpha$ goes to zero due to the square-root dependence
on the coefficient of restitution in Eq.(\ref{eq:32}).
\begin{figure}[t]
\begin{center}
\resizebox{11cm}{!}{\includegraphics{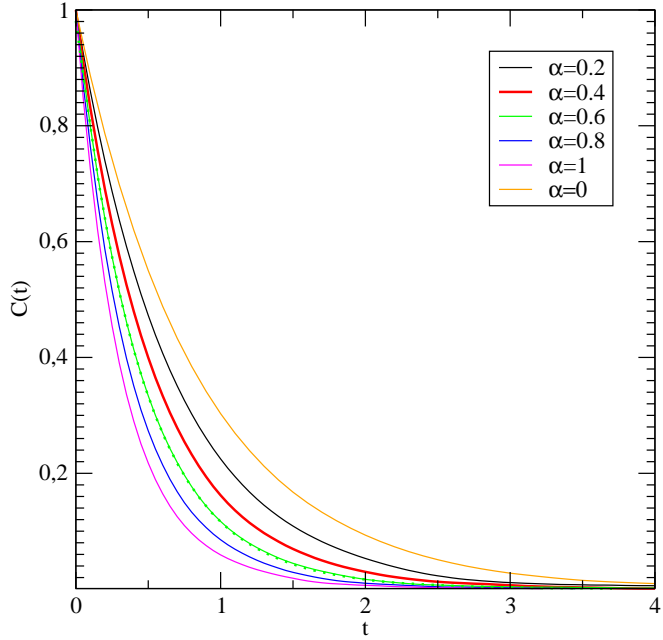}}

\caption{Normalized autocorrelation function $C(t)$ of the velocity 
versus time for $M=0.6m$ and for  various values of the coefficient of
restitution $\alpha=1.0,0.8,0.6,0.4,0.2,0$. The dashed curve corresponds to
the inverse Laplace transform of Eq.(\ref{eq:30})}\label{fig:3}
\end{center}
\end{figure}
\section{Simulation results}  In order  to study the  situations where
$M/m\neq\alpha$, we have  solved the Boltzmann equation numerically by
using  a  DSMC method \cite{MS00}.    We focus here on  the stationary
dynamics.  Figure  \ref{fig:3} displays both  the numerical results of
the  simulation and the  exact results obtained  by an inverse Laplace
transform of Eq.(\ref{eq:30}): the  ratio of the tracer  particle over
the mass of     the bath  particle    is  equal  to  $\alpha$,    when
$\alpha=0.6$. Note that   decreasing the  coefficient  of  restitution
increases    the  relaxation     time of     the   normalized velocity
autocorrelation  function   but does   not  significantly  change  its
shape.  The  dashed   curve   corresponds   to the   exact   solution,
Eq. (\ref{eq:30}), and  matches very accurately the simulation results
when $\alpha=0.6$.

As $\alpha$ decreases, the velocity autocorrelation function decreases
more   slowly. This behavior   is,  at first glance, counterintuitive.
When  $\alpha$ is small the postcollisional  velocity of the tracer is
smaller than  when $\alpha$ is larger. One  therefore expects that the
correlation function should approach zero  more rapidly when  $\alpha$
is small.  This reasoning,  however,  does not  account for  the  time
between collisions which increases as $\alpha$ decreases. Evidently it
is this effect that dominates, leading to the observed behavior.

\section{Self correlation function}
The more general case   of spatially inhomogeneous  states
also admits an analytic  solution of the Boltzmann equation.  
Since the strategy is basically similar   to that developed    in  Section 3, details of
calculations are given in Appendix A. The  final result for $\tilde{f}(q,u,z)$ reads
\begin{align}
\label{eq:34}
\tilde{f}(q,u,z)&=\phi(u)\left[\frac{1}{z}-\frac{1}{z+\nu(u)+iqu}+\right.\nonumber\\
&\left.\int_{-\infty}^udw\frac{1}{[z+\nu(w)+iqw]^2}
\left(\frac{\partial \hat{G}}{\partial w}(q,w,0)-\frac{d\nu}{d w}(w)+<\tilde{v}(q,z)>\right)\right]
\nonumber\\&+\frac{1}{z+\nu(u)+iqu}f(q,u,0).
\end{align}
Here $\hat{G}$ denotes the Fourier transform
\begin{equation}
\label{Fourier}
 \hat{G}(q,u,0) = \int_{-\infty }^{+\infty }dre^{iqr}G(r,u,0) 
\end{equation}
and $G(r,u,t)$ is defined by Eq.(\ref{eq:A1}) in analogy with the definition (\ref{eq:12}).

The conditional density $F(r,t)$ can be  obtained  from the inhomogeneous  Boltzmann
equation by integrating $f(r,v,t)$ over   the velocities assuming  that  
\begin{equation}
\label{initial}
f(r,v,0)=\delta(r)\phi_M(u)
\end{equation}
 i.e.   that initially the tracer 
particle was  located at the origin with the equilibrium velocity distribution.  This is 
in fact the definition of the  self-correlation function
\begin{equation}
F(r,t)=\int_{-\infty}^{+\infty}dv f(r,v,t).
\end{equation}
As, in accordance with the definition (\ref{eq:A1}),
\begin{equation}
 2f(r,v,t) = \frac{\partial^2}{\partial v^{2}}G(r,v,t) 
\end{equation}
one finds that  the Fourier-Laplace transform of $F(r,t)$ is given by
\begin{equation}
\tilde{F}(q,z)=\int dr\, e^{-iqr}\int_{0}^{\infty}dt\, e^{-zt}F(r,t)=
\frac{1}{2}\left[\frac{\partial\tilde{G}}{\partial v}(q,+\infty,z)-\frac{\partial\tilde{G}}{\partial v}(q,-\infty,z)\right]
\end{equation}
From Eq.(\ref{eq:A11})  a simple expression follows
\begin{equation}
\tilde{F}(q,z)=\frac{1-iq<\tilde{v}(q,z)>}{z},
\end{equation}
whereas using the initial condition Eq.(\ref{initial}) one finds the formula
\begin{equation}
<\tilde{v}(q,z)>=iq\left[ \frac{z(1+q^2)J(q,z)-2}{z(1+q^2)J(q,z)+2q^2}\right]
\end{equation}
with 
\begin{equation}\label{eq:35}
\tilde{J}(q,z)=\int_{-\infty}^{+\infty}\frac{dw}{[z+iqw+\nu(w)]^2}
\end{equation}
Finally, one gets
\begin{equation}
\tilde{F}(q,z)=\frac{1}{z}+\frac{q^2}{z}\left(1-\frac{2(1+q^2)}{z(1+q^2)\tilde{J}(q,z)+2q^2}\right)
\end{equation}

The wave vector dependent poles of $\tilde{F}(q,z)$ satisfy the equation
\begin{equation}\label{eq:36}
z(1+q^2)\tilde{J}(q,z)+2q^2=0
\end{equation}
For the hydrodynamic pole $z(q)$, which tends to zero when $q\to 0$,
  Eq.(\ref{eq:36}) to dominant order  yields
\begin{align}
z\tilde{J}(0,0)+2q^2=0\\
z+Dq^2=0
\label{eq:37}
\end{align}
where     $D$  is     the  diffusion  coefficient 
derived  in section~\ref{sec:corr-funct-transp} with the use of the 
Green-Kubo autocorrelation formula (see Eq.(\ref{eq:31}) ). Here it
appears in the hydrodynamic long-time and long-length scale
limit of the solution of the Boltzmann equation which expresses the density fluctuation of the tracer particle.

\section{Conclusion}
We have shown
that, within the framework of the one-dimensional Boltzmann equation,
the dynamics of a granular tracer particle can be mapped onto the
dynamics of an elastic tracer particle with a renormalized mass. A
recent article of Santos and Dufty \cite{SD06} shows that this result can
be extended to more general situations.

J. P. acknowledges  financial support by the CNRS, France, and
the hospitality at the {\it Laboratoire de Physique Th\'eorique de la Mati\`ere
Condens\'ee, UPMC  (Paris)} where this research
has been carried out.

\appendix

\section{Inhomogeneous solutions of the Boltzmann equation}
We summarize the basic steps leading to the solution of the
inhomogeneous   Boltzmann equation.   The quantity   which
generalizes Eq.(\ref{eq:12}) is the function $G(r,u,t)$ defined as
\begin{equation}
\label{eq:A1}
G(r,u,t)=\int_{-\infty }^{+\infty }dc|u-c|f(r,c,t).
\end{equation}
Let us denote $\tilde{G}(q,u,z)$ the Fourier-Laplace transform of $G(r,u,t)$, 
\begin{equation}
\label{eq:A2}
\tilde{G}(q,u,z)=\int_{-\infty }^{+\infty }dre^{iqr}\int_0^{+\infty }dt e^{-zt}G(r,u,t).
\end{equation}
The Boltzmann equation, Eq.(\ref{eq:5}), can be then rewritten as
\begin{equation}
\label{eq:A3}
[z+iqu+\nu(u)]\frac{\partial^2 \tilde{G}}{\partial  u^2}(q,u,z)-\frac{d^2\nu (u)}{d  u^2}
\tilde{G}(q,u,z)=\frac{\partial^2 \hat{G}}{\partial u^2}(q,u,0)
\end{equation}
Integrating Eq.(\ref{eq:A3}) over the velocity interval $]-\infty, u]$ yields
\begin{equation}
\label{eq:A4}
[z+iqu+\nu(u)]\frac{\partial \tilde{G}}{\partial  u}(q,u,z)-\left[\frac{\partial\nu (u)}{\partial  u}+iq\right]
\tilde{G}(q,u,z)=\frac{\partial \hat{G}}{\partial u}(q,u,0)+\tilde{B}(q,z)
\end{equation}
where the function $\tilde{B}(q,z)$ is to be determined.
Similarly to Eq.(\ref{eq:20}), it is convenient to introduce the function $\tilde{H}(q,u,z)$ 
\begin{equation}
\tilde{H}(q,u,z)=\frac{\tilde{G}(q,u,z)}{[z+iqu+\nu(u)]},
\end{equation}
which satisfies  a  differential equation analogous to  Eq.(\ref{eq:21})
with $\nu(u)$  replaced  by  $[\nu(u)+iqu]$, and $<\tilde{v}(z)>$ replaced by $<\tilde{v}(q,z)>$. 
Therefore, the solution  of Eq.(\ref{eq:A3}) can be expressed as
\begin{equation}
\label{eq:A5}
\tilde{G}(q,u,z)=[z+\nu(u)+iqu]
\end{equation}
\[\times \left[\int_{-\infty}^udw\frac{1}{[z+\nu(w)+iqw]^2}
\left( \frac{\partial \hat{G}}{\partial w}(q,w,0)+\tilde{B}(q,z)\right)+D(q,z)\right]  \]

The functions $B(q,z)$  and   $D(q,z)$  can be determined    from  the
boundary conditions satisfied by $\tilde{G}(q,u,z)$ and $\tilde{H}(q,u,z)$.
The asymptotic behavior of $\tilde{G}(q,u,z)$ for large $u$ at fixed $z$ and $q$ is given by
\begin{equation}
\label{eq:A6}
\tilde{G}(q,u,z)\sim |u|\int_{-\infty}^{+\infty}dw\tilde{f}(q,w,z)-sgn(u)<\tilde{v}(q,z)>
\end{equation}
where
\begin{equation}\label{eq:A7}
<\tilde{v}(q,z)>=\int_{-\infty}^{+\infty}dww\tilde{f}(q,w,z)
\end{equation}
The continuity equation (conservation of the probability) is expressed as
\begin{equation}
\label{eq:A8}
z\int_{-\infty}^{+\infty}dw\tilde{f}(q,w,z)+iq<\tilde{v}(q,z)>=1
\end{equation}
By combining Eqs.(\ref{eq:A6}) and (\ref{eq:A8}), one obtains the asymptotic relation
\begin{equation}
\label{eq:A9}
\tilde{G}(q,u,z) \sim \frac{|u|}{z}[1-iq<\tilde{v}(q,z)>]-sgn(u)<\tilde{v}(q,z)>
\end{equation}

The limit  of $\tilde{H}(q,u,z)$ when $u\to\pm\infty $ is then given by
\begin{equation}
\label{eq:A99}
\lim_{u\to\pm\infty }\tilde{H}(q,u,z)=\frac{1-iq<\tilde{v}(q,z)>}{(1\pm iq)z}
\end{equation}
By  integrating  the differential equation obeyed by
$\tilde{H}(q,u,z)$ over  velocity, one obtains a sum rule analogous to Eq.(\ref{eq:23}):
\begin{equation}
\label{eq:A10}
<\tilde{v}(q,z)>=-\left(\frac{2iq}{1+q^2}+\int_{-\infty}^{+\infty}\frac{du}{[z+\nu(u)+iqu]^2}
\frac{\partial \hat{G}}{\partial u}(q,u,0)\right)
\end{equation}
\[ \times \left( \int_{-\infty}^{+\infty}\frac{du}{[z+\nu(u)+iqu]^2}+\frac{2q^2}{1+q^2}\right)^{-1} \]
The knowledge of the boundary conditions (\ref{eq:A9}) and (\ref{eq:A99})
 permits one to write the solution (\ref{eq:A5}) in an explicit form 
\begin{equation}
\label{eq:A11}
\tilde{G}(q,u,z)=[z+\nu(u)+iqu]
\end{equation}
\[ \times \left[\int_{-\infty}^udw\frac{1}{[z+\nu(w)+iqw]^2}
\left(\frac{\partial \hat{G}}{\partial w}(q,w,0)+<\tilde{v}(q,z)>\right)+\frac{1-iq<\tilde{v}(q,z)>}{(1- iq)z}\right] \]
Taking the second derivative of Eq.(\ref{eq:A11}) with respect to the velocity variable $u$ 
leads to Eq.(\ref{eq:34}).


\begin{thebibliography}{10}
\bibitem{exactptv}
Note that for one-dimensional dilute gas, the particule size is irrelevant.
\end{thebibliography}

\end{document}